\documentclass[prb,twocolumn,msmath,superscriptaddress,amssymb,showpacs]{revtex4}

\usepackage{color} 
\usepackage{graphicx}
\usepackage{epstopdf}
\usepackage{bm}

\begin{document}

\DeclareGraphicsExtensions{.eps, .png, .jpg}
\bibliographystyle{prsty}

\title {Electrodynamic properties of an artificial heterostructured superconducting cuprate}

\author{A. Perucchi}
\affiliation{Elettra - Sincrotrone Trieste, S.S. 14 km 163.5 in AREA SCIENCE PARK, 34012 Basovizza, Trieste, Italy}
\author{P. Di Pietro}
\affiliation{Elettra - Sincrotrone Trieste, S.S. 14 km 163.5 in AREA SCIENCE PARK, 34012 Basovizza, Trieste, Italy}
\author{S. Lupi}
\affiliation{CNR-IOM and Dipartimento di Fisica, Universit\`a di Roma ÒSapienzaÓ, Piazzale A. Moro 2, 00185 Roma, Italy}
\author{R. Sopracase}
\affiliation{GREMAN, CNRS UMR 7347-CEA, UniversitŽ F. Rabelais}
\author{A. Tebano}
\affiliation{CNR-SPIN and Dipartimento di Ingegneria Civile e Ingegneria Informatica, Universit\`a di Roma Tor Vergata, Via del Politecnico 1, 00133- Roma, Italy}

\author{G. Giovannetti}
\affiliation{International School for Advanced Studies (SISSA) and CNR-IOM-Democritos National Simulation Centre, Via Bonomea 265, I-34136 Trieste, Italy}
\affiliation{Institute for Theoretical Solid State Physics, IFW-Dresden, PF 270116, 01171 Dresden, Germany}

\author{F. Petocchi}
\affiliation{International School for Advanced Studies (SISSA) and CNR-IOM-Democritos National Simulation Centre, Via Bonomea 265, I-34136 Trieste, Italy}

\author{M. Capone}
\affiliation{International School for Advanced Studies (SISSA) and CNR-IOM-Democritos National Simulation Centre, Via Bonomea 265, I-34136 Trieste, Italy}
\author{D. Di Castro}
\affiliation{CNR-SPIN and Dipartimento di Ingegneria Civile e Ingegneria Informatica, Universit\`a di Roma Tor Vergata, Via del Politecnico 1, 00133- Roma, Italy}

\date{\today}

\begin{abstract}
We perform infrared conductivity measurements on a series of CaCuO$_2$/SrTiO$_3$ heterostructures made by the insulating cuprate CaCuO$_2$ (CCO) and the insulating perovkite SrTiO$_3$ (STO). We estimate the carrier density of various heterostructures, with different level of hole doping from the integral of the optical conductivity and we measure the  corresponding degree of correlation by estimating the ratio between the Drude weight and the integral of the infrared spectrum. The analysis demonstrates a large degree of correlation, which increases as the doping is reduced. The experimental results can be reproduced by Dynamical Mean-Field Theory calculations which strongly support the role of correlations in the CCO/STO heterostructures and their similarities with the most common cuprate superconductors. Our results suggest that cuprate superconductors can be looked at as natural superlattices, where the properties of the CuO$_2$ conducting planes and charge reservoir blocks can be completely disentangled.
\end{abstract}

\maketitle

The progress in the ability to grow oxide heterostructures with atomically sharp interfaces has lead to the discovery of a wealth of fascinating states of matter found at the interface between two distinct materials. Two-dimensional metallic and superconducting states,  magnetism, quantum hall effect, are few spectacular examples of the phenomena \cite{hwang12} that can nowadays be engineered with layer by layer growth techniques, thus allowing to achieve unprecedented control on materials' functionalities. 

Atomic scale engineering techniques can also be used to control widely known phenomena in well established classes of transition-metal oxides. A notable example is realized with manganites, where the  LaMnO$_3$/SrMnO$_3$ heterostructures \cite{perucchi10} allow to address the physics of double exchange in the absence of substitutional disorder which is unavoidable when doping is realized via alloying. 

A similar approach can be adopted to tackle the three-decades old problem of high temperature superconductivity in the cuprates. Here, the key idea is to disarticulate the cuprate structure in terms of CaCuO$_2$ "infinite layers", which provide the crucial CuO$_2$  planes, and of a perovskite (copper free) block acting as a charge reservoir. The CCO/STO superlattice has been recently explored, providing evidence for high temperature superconductivity up to 50 K \cite{dicastro12,dicastro14}. In this artificial structure one can decide the number $n$ of adjacent CuO$_2$  planes, while controlling through $m$ (number of SrTiO$_3$ unit cells) the distance between the CaCuO$_2$ blocks. As shown in Ref.\onlinecite{dicastro12}, this enables to tune the superconducting properties and find the optimal structure. Doping is controlled by varying the oxidization conditions during the growth. Oxygen ions diffuse in the spaces made available at the interface  between the infinite layer and the perovskite structure, penetrating into the Ca planes of the CaCuO$_2$ layers \cite{dicastro15}.

These artificial superconductors have been characterized before through transport and X-ray absorption spectroscopy\cite{dicastro12, dicastro14}, magnetotransport (anisotropy)\cite{salvato13}, Raman\cite{dicastro13b}, RIXS\cite{minola12},  HAXPES\cite{aruta13} and STEM/EELS\cite{dicastro15}. In this work we employ infrared spectroscopy to explore the electrodynamic properties of the CCO/STO superlattices at various doping levels. Our analysis clearly shows characteristic fingerprints of strong electronic correlations, similar to more standard superconducting cuprates as YBCO or LSCO. The experimental study is supported by theoretical calculations based on Dynamical Mean Field Theory (DMFT)\cite{dmft}, which allows for a more quantitative assessment of the degree of correlation through a direct comparison of the optical spectra and the associated sum rules\cite{toschi05,toschiPRB,basov11,Nicoletti}. 

Four samples of CCO/STO have been prepared following the procedures described in Ref. \onlinecite{dicastro12}. The four samples (SL$\#1$ to SL$\#4$) are made up of equal numbers of CaCuO$_3$ (CCO) and SrTiO$_3$ (STO) unit cells, i.e. $n=m=3$, but with increasing doping level due to the different oxydizing atmosphere used during the growth (oxygen pressure varies from 0.2 mbar to 0.8 mbar). The first three samples (SL$\#1$ to SL$\#3$) do not display superconductivity down to at least 15 K. The fourth sample (SL$\#4$) has a higher doping level (obtained by growing the SL in a mixture of oxygen and 12\% ozone at 0.8 mbar), resulting in a superconducting behavior with  $T_c$ = 20.5 K, as shown by resistivity measurements (see inset of Fig.4). All the samples are grown on (La,Sr)(Al,Ti)O$_3$ (LSAT) substrates. The films have a thickness between 40 and 50 nm. One single CCO film on LSAT has been measured as well, for reference. The reflectivity measurements were performed at the IR beamline SISSI at the Elettra synchrotron\cite{lupi07}, at nearly normal incidence with the help of a Bruker 70v spectrometer equipped with sources, beamsplitters, and detectors able to cover the whole infrared range up to the visible.

\begin{figure}[t]
\begin{center}
\leavevmode
\includegraphics [width=8cm]{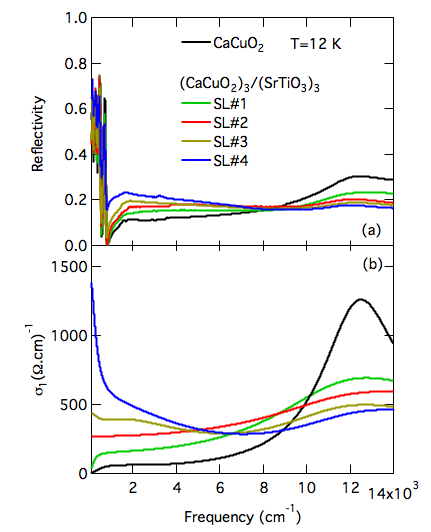}  
\end{center}
\caption{Reflectivity (a) and real part of the optical conductivity (b) of the 4 CCO/STO superlattices, and of a CCO film. The optical conductivity clearly shows the progressive depletion of the absorption peak at 12000 cm$^{-1}$, and the corresponding accumulation of spectral weight at infrared frequencies }
\label{Fig1}
\end{figure}

\begin{figure}[t]
\begin{center}
\leavevmode
\includegraphics [width=8cm]{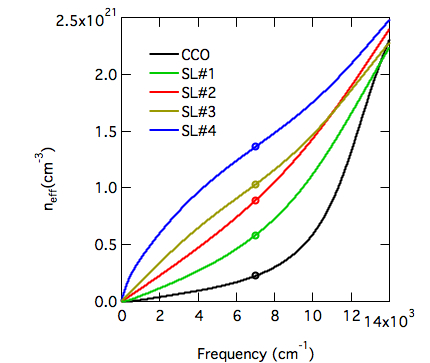}  
\end{center}
\caption{Frequency dependent spectral weight. The effective number of carriers is calculated according to equation (\ref{neff}), by assuming a cut-off frequency value at 7000 cm$ö{-1}$ (see text). }
\label{Fig2}
\end{figure}

Figure \ref{Fig1} summarizes the main results of the present work. The upper panel reports the raw reflectivity data at 12 K which were fitted with the help of a phenomenological, Kramers-Kronig consistent, Drude-Lorentz model \cite{Dressel, kuzmenko05,perucchi10}. From the fit parameters it is possible to calculate the optical conductivity which is shown in the lower panel, after subtracting the phonon modes. The optical conductivity of CCO is characterized by one single absorption centered at about 12000 cm$^{-1}$ (1.5 eV), which we attribute to a charge transfer band. With increasing doping a broad absorption in the infrared range appears, gradually evolving into a Drude-like peak. Such increased absorption grows at the expenses of the charge transfer band. This evolution of the frequency dependence of the optical spectra is summarized in Fig. \ref{Fig2}, where we plot the integral of the spectral weight as a function of an integration cut-off $\Omega$. 
According to the f-sumrule, we can define an effective carrier density as:
\begin{equation}
n_{eff}(\Omega)=\frac{2m_e}{\pi e^2}\int_0^{\Omega}\sigma_1(\omega)d\omega, \label{neff}
\end{equation}
where $e$ and $m_e$ are the bare electron charge and mass. 

In the absence of direct quantitative  information on the doping level, we employ the optical measurement to extract the carrier density of our superlattice compounds. As previously shown in Ref.\onlinecite{padilla05}, the total spectral weight calculated below the charge transfer absorption gap works as a reasonable proxy of the charge density, over a large region of the cuprate phase diagram. In order to calculate this quantity we extract the $n_{eff}$ value calculated at a cut-off frequency of 7000 cm$^{-1}$, which corresponds to a minimum in the optical conductivity for the two more doped superlattices SL$\#3$ and SL$\#4$. This choice is meant to separate the conductivity due to the doped carriers from that related to interband transitions. It should be noted that this choice is not unique and it has been the subject of debate in the past\cite{carbone06}. An incertitude of about $\pm$ 1000 cm$^{-1}$ in the cut-off frequency leads to an error bar of $\pm$ 10\% in the determination of the carrier density. 

By combining structural data\cite{dicastro15} with the carrier density extracted as discussed above, we can finally establish the number of charge carriers (holes) per Cu ion. This results in hole doping values ranging from 0.05 for the most insulating compound (SL$\#1$) to 0.14 for the superconducting sample SL$\#4$. We note that these values are typical for the underdoped regime of most common cuprate compounds\cite{basov05}. 

As it is clearly seen in the case of the SL$\#4$ compound, the in-gap infrared spectral weight associated to charge carriers can be roughly divided in two parts: a Drude term with a linewidth of about 700 cm$^{-1}$, and a broad absorption band extending over the whole mid-infrared range. 
The existence of these two features can be understood in rather general terms as a consequence of strong correlations. In a strongly correlated system, the electron behaves like an itinerant quasiparticle at low-energy and as an almost localized object at high energy. The Drude contribution arises from optical transitions involving the itinerant states, while the mid-infrared feature is associated to processes connecting the incoherent lower Hubbard band and the quasiparticle peak \cite{rozenberg95,toschi05, baldassarre08,lupi04}. When correlations increase, the spectral weight shifts from the Drude peak to the mid-infrared response as a consequence of the reduced mobility of the carriers due to the approach to a Mott transition.

This suggests that the ratio between the Drude spectral weight and the full mid-infrared spectral weight  provides a mean to estimate the degree of correlation in our compounds which is based only on measured quantities\cite{lovecchio15}. The smaller this ratio, the larger the correlation effects reducing the weight of the Drude contribution. This quantity conveys a similar information as the ratio between the experimental kinetic integrated spectral weight (which can be connected with the kinetic energy for a lattice model) and the non-interacting kinetic energy which can be estimated by means of theoretical density-functional calculations in which the electron-electron correlations are not included \cite{qazilbash09,degiorgi11}. 

\begin{figure}[t]
\begin{center}
\leavevmode
\includegraphics [width=8cm]{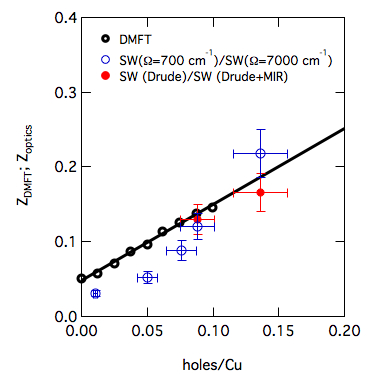}  
\end{center}
\caption{Evaluation of the quasi-particle spectral weight $Z$ from optics (blue and red circles) , according to spectral weight arguments (see text) and DMFT (black circles). }
\label{Fig3}
\end{figure}

In this work we use two different estimates of the Drude and mid infrared spectral weights. The first choice is to use the spectral weight components obtained from a Lorentz-Drude fitting model ($Z_{optics}=\frac{SW_{Drude}}{SW_{Drude}+SW_{MIR}}$, where $SW_{Drude/MIR}=\int\sigma_{Drude/MIR}(\omega) d\omega$), while the second estimate is given by direct integration of the spectral weight with two different cut-offs. In the latter case we integrate up to $\Omega$=700 cm$^{-1}$ (corresponding to the onset of the Drude peak) to  estimate the Drude weight, and up to $\Omega$=7000 cm$^{-1}$ (the cut-off frequency already chosen to identify the in-gap spectral weight) in order to gauge the  mid-infrared weight ($Z_{optics}=\frac{SW(\Omega=700 cm^{-1})}{SW(\Omega=7000 cm^{-1})}$, where $SW(\Omega)=\int_0^{\Omega}\sigma(\omega) d\omega$).
The results are reported in Fig. \ref{Fig3} as a function of the estimated doping level. We note here that in the case of the values based on the Lorentz-Drude fit, we can define the $\frac{SW_{Drude}}{SW_{Drude}+SW_{MIR}}$ ratio only for the two most metallic samples (SL$\#3$ and SL$\#4$), since for the other two samples the Drude term can not be properly defined. The two estimates are compatible and they both show a relatively high degree of correlation, as signalled by the small values of the $\frac{SW_{Drude}}{SW_{Drude}+SW_{MIR}}$ ratio (comparable to what found in copper oxide superconductors\cite{lovecchio15}), and a monotonic decrease of correlations (increase of the ratio) as the doping grows.

This purely experimental analysis clearly shows a substantial strength of correlations and a doping dependence fully corresponding with a degree of correlation which is mainly determined by the distance from a half-filled band, as expected in a doped Mott insulator. In order to give a more quantitative account for these
results and to put this observation on more solid ground we have performed Dynamical Mean-Field Theory (DMFT) calculations
starting from a density-functional theory (DFT) ab-initio description of a supercell composed by three CCO and three STO layers. The effective impurity model is solved using exact diagonalization as detailed in Ref. \onlinecite{ED}.
We relaxed atomic positions within Local Density Approximation (LDA) starting from the experimental lattice spacing of 3.87 $\AA$  in the plane \cite{nota}
and a perpendicular length given by three bulk CCO and three bulk STO unit cells. 
 In the calculations we have used VASP\cite{VASP} 
with a 500 eV cut-off and a 12$\times$12$\times$2 k-point grid. After the relaxation the Ti atoms are off-centered.

Within LDA we find that the titanium d-orbitals of the STO component are essentially empty,
while the copper-oxygen CCO states at the Fermi level have mainly d$_{x2-y2}$ character. Therefore we built a single-orbital representation of the electronic
structure using Wannier90\cite{wannier90}  to build maximally localized Wannier orbitals. The valence band can be well reproduced by retaining only the d$_{x2-y2}$--like orbital. This gives rise to a single-band model completely analogous to the standard cuprates. The metallic character observed in LDA is obviously a consequence of the neglect of Coulomb interaction which is notoriously important for copper oxides and in particular for the cuprates. 

In the absence of an ab-initio estimate of the Coulomb interaction appropriate for our heterostructure,
we use a value of $U = 4.9$ eV, which is chosen to provide a satisfactory agreement with experiments. The qualitative 
trends of the calculations do not depend on the choice of the interaction as long as it is sufficiently strong to induce
sizable correlation effects.

In Fig. 3 we compare the experimental estimate of correlations with the quasiparticle weight $Z$, which measures
the fraction of the total optical weight belonging to the Drude peak. Our calculations clearly reproduce nicely the doping dependence of the experimental 
estimates. $Z$ varies only slightly in the three correlated CCO
layers and we plot the average. Interestingly the inclusion of correlations reduces the differences between the three CCO layers with respect to the LDA
calculations.

\begin{figure}
\includegraphics[width=8cm]{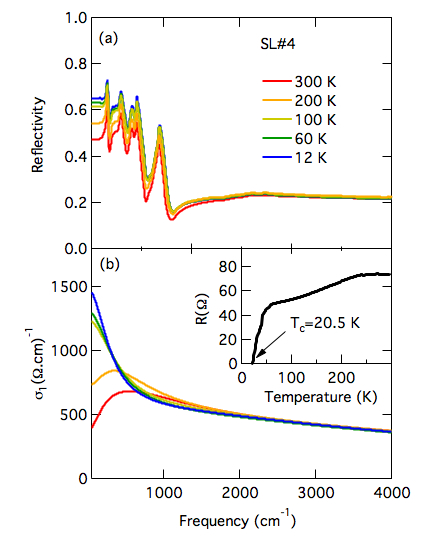}
\vspace{0.25cm}
\caption{Temperature-dependent reflectivity (a) and real part of the optical conductivity (b) for the superconducting SL$\#4$  CCO/STO superlattice. The inset displays the temperature-dependent resistance of the film.}
\label{Fig4}
\end{figure}

The role of electronic correlations in shaping the optical properties of the CCO/STO superlattices can be further investigated by analysing the temperature dependence of the  electrodynamic response. We focus here on the SL$\#4$ compound, which is the sample displaying the largest temperature dependence (see Fig. \ref{Fig4}). Significant variations in the optical conductivity as a function of temperature are observed below 2000 cm$^{-1}$ only. In particular, upon increasing temperature one observes a depletion of the Drude peak, which disappears between 100 and 200 K. At room temperature the Drude peak is completely washed out by thermal fluctuations and the optical conductivity shows a maximum at a finite frequency of about 500 cm$^{-1}$, as an evidence for a significant loss of coherence at high temperatures. Interestingly, above 200 K, the resistivity of the SL$\#4$ compound also flattens (see inset of Fig. \ref{Fig4}), thus suggesting that this temperature may coincide with the onset of a pseudogap, as observed in underdoped cuprates. 

\begin{figure}
\includegraphics[width=8cm]{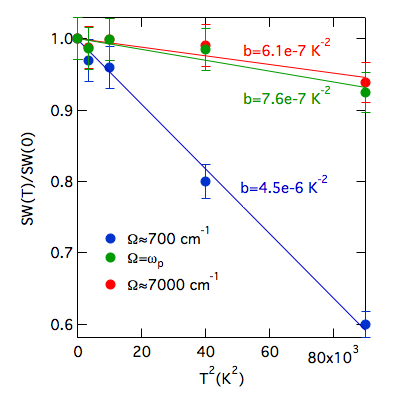}
\vspace{0.25cm}
\caption{Temperature-dependence of the partial spectral weight (SW) for different cut-off frequencies for the  superconducting SL$\#4$ CCO/STO superlattice. The spectral weight dependence is compatible with the predicted $1-b(\Omega)T^2$ behavior (see Ref. \onlinecite{toschi05}) with slope comparable to what found in other cuprates\cite{ortolani05,lovecchio15}.}
\label{Fig5}
\end{figure}

Finally, we analyse the temperature dependence of the spectral weight. According to Refs \onlinecite{ortolani05,toschi05}, within a tight-binding framework, the partial spectral weight $SW(\Omega)$ is a quantity directly related to the kinetic energy, and therefore displays a temperature dependence in the form 
\begin{equation}
\frac{SW(\Omega,T)}{SW(\Omega,0)}\simeq 1-b(\Omega)T^2
\end{equation}
In a conventional metal, one expects the $b(\Omega)$ value to be zero for $\Omega\geqslant\omega_p$, where $\omega_p$ is the plasma frequency. On the other hand in correlated materials the $b(\Omega=\omega_p)$ coefficient is sizable at the plasma frequency, in the range of 10$^{-7}$K$^{-2}$.  We report in Fig. \ref{Fig5}, the $T^2$ plot of the spectral weight, for several cut-off values extending from $\Omega$=700 cm$^{-1}$ to 7000 cm$^{-1}$. The value of the $b(\Omega=\omega_p)$ coefficient is 7.6$\times$10$^{-7}$K$^{-2}$, comparable to what found previously in other copper oxide superconductors\cite{ortolani05,lovecchio15}. 

Our infrared measurements and the related DMFT calculations clearly show that the electrodynamics 
of artificial superconducting CCO/STO superlattices and conventional superconducting cuprates share many similarities and has a similar degree of correlation.
Therefore we can gain information on the elusive properties of the cuprates by comparing with our controllable artificial systems. 
 Indeed, in CCO/STO the conductivity (and thus the superconductivity) is a pure interfacial phenomenon: the holes are injected in the cuprate from the interface Ca plane and extend within the cuprate only for 1-2 unit cells \cite{dicastro15}. This leads to the appealing idea that the conventional cuprates can be considered as natural superlattices, where the superconducting and electrodynamic properties are dominated by the physics of the native interfaces between the conducting block (with the CuO$_2$ planes) and the charge reservoir block. 

\section*{Acknowledgements}
MC and FP acknowledge financial support from MIUR through the PRIN 2015 program (Prot. 2015C5SEJJ001) and SISSA/CNR project "Superconductivity, Ferroelectricity and Magnetism in bad metals" (Prot. 232/2015)



\begin{thebibliography}{30}
\bibitem{hwang12} H.Y. Hwang, Y. Iwasa, M. Kawasaki, B. Keimer, N. Nagaosa and Y. Tokura, Nat. Materials {	\bf 11}, 103 (2012)

\bibitem{perucchi10} A. Perucchi, L. Baldassarre, A. Nucara, P. Calvani, C. Adamo, D.G. Schlom, P. Orgiani, L. Maritato and S. Lupi, Nano Letters {\bf 10}, 4819 (2010)

\bibitem{dicastro12} D. Di Castro {\it et al.}, Phys. Rev. B {\bf 86}, 134524 (2012)

\bibitem{dicastro14} D. Di Castro, C. Aruta, A. Tebano, D. Innocenti, M. Minola, M.M. Sala, W. Prellier, O. Lebedev and G. Balestrino, Supercond. Sci. Technol. {\bf 27}, 044016 (2014)

\bibitem{dicastro15} D. Di Castro, C. Cantoni, F. Ridolfi, C. Aruta, A. Tebano, N. Yang and G. Balestrino, Phys. Rev. Lett. {\bf 115}, 147001 (2015)

\bibitem{salvato13} M Salvato, I Ottaviani, M Lucci, M Cirillo, D Di Castro, D Innocenti, A Tebano, G Balestrino, J. Phys.: Condens. Matter 25 (2013) 335702

\bibitem{dicastro13b} D. Di Castro, S. Caramazza, D. Innocenti, G. Balestrino, C. Marini, P. Dore, and P. Postorino, Appl. Phys. Lett. 103, 191903 (2013)

\bibitem{minola12} M. Minola, D. Di Castro, L. Braicovich, N.B. Brookes, D.
Innocenti, M. Moretti Sala, A. Tebano, G. Balestrino, and
G. Ghiringhelli, Phys. Rev. B 85, 235138 (2012).

\bibitem{aruta13} C. Aruta, C. Schlueter, T.-L. Lee, D. Di Castro, D. Inno-
centi, A. Tebano, J. Zegenhagen, and G. Balestrino, Phys.
Rev. B 87, 155145 (2013).

\bibitem{dmft} A. Georges, G. Kotliar, W. Krauth, and M
.J. Rozenberg, Rev. Mod. Phys. {\bf 68}, 13 (1996)

\bibitem{toschi05} A. Toschi {\it et al.}, Phys. Rev. Lett. {\bf 95}, 097002 (2005).

\bibitem{toschiPRB} A. Toschi and M. Capone,  Phys. Rev. B {\bf 77}, 014518 (2008)

\bibitem{basov11} D. N. Basov, Richard D. Averitt, D. van der Marel, M. Dressel, and K. Haule, Rev. Mod. Phys. {\bf 83}, 471 (2011)

\bibitem{Nicoletti} D. Nicoletti, O. Limaj, P. Calvani, G. Rohringer, A. Toschi, G. Sangiovanni, M. Capone, K. Held, S. Ono, Yoichi Ando, and S. Lupi
Phys. Rev. Lett. 105, 077002 (2010)

\bibitem{lupi07} S. Lupi, A. Nucara, A. Perucchi, P. Calvani, M. Ortolani, L. Quaroni, M. Kiskinova, J. Opt. Soc. Am. B {\bf 24}, 959 (2007)

\bibitem{Dressel}M. Dressel and G. Gr\"uner, Electrodynamics of Solids: optical properties of electrons in matter, Cambridge University Press (2002).

\bibitem{kuzmenko05} A.B. Kuzmenko, Rev. Sci. Instrum. {\bf 76}, 083108 (2005).

\bibitem{padilla05} W.J. Padilla, Y.S. Lee, M. Dumm, G. Blumberg, S. Ono, Kouji Segawa, Seiki Komiya, Yoichi Ando, and D.N. Basov, Phys. Rev. B {\bf 72}, 060511 (2005).

\bibitem{carbone06} F. Carbone, A.B. Kuzmenko, H.J.A. Molegraaf, E. van Heumen, E. Giannini, and D. van der Marel, Phys. Rev. B {\bf 74}, 024502 (2006).

\bibitem{basov05} D. Basov and T. Timusk, Rev. Mod. Phys. {\bf 77}, 721 (2005).

\bibitem{rozenberg95} M.J. Rozenberg, G. Kotliar, H. Kajueter, G.A. Thomas, D.H. Rapkne, J.M. Honig, and P. Metcalf, Phys. Rev. Lett. Phys. {\bf 75}, 105 (1995).

\bibitem{baldassarre08} L. Baldassarre, A. Perucchi, D. Nicoletti, A. Toschi, G. Sangiovanni, K. Held, M. Capone, L. Malavasi, M. Marsi, P. Metcalf, P. Postorino, and S. Lupi, Phys. Rev. B {\bf 77}, 113107 (2008).

\bibitem{lupi04} S. Lupi, M. Ortolani, and P. Calvani, Phys. Rev. B {\bf 69}, 180506 (2004).

\bibitem{lovecchio15} I. Lo Vecchio, L. Baldassarre, F. D'Apuzzo, O. Limaj, D. Nicoletti, L. Fan, P. Metcalf, M. Marsi, and S. Lupi, Phys. Rev. B {\bf 91}, 155133 (2015).

\bibitem{qazilbash09} M.M. Qazilbash {\it et al.}, Nature Physics 5, {\bf 647} (2009).

\bibitem{degiorgi11} L. Degiorgi, New J. Phys. {\bf 13}, 023011 (2011).

\bibitem{ED} M. Capone, L. de' Medici, and A. Georges, Phys. Rev. B {\bf 76}, 245116 (2007)

\bibitem{nota} This is the value of the LSAT substrate lattice parameter. Indeed, it has been shown in Ref. \onlinecite{dicastro14} that in CCO/STO SLs CCO and STO are strained within the in-plane direction and adopt the in-plane lattice paramters of the substrate.

\bibitem{VASP} G. Kresse {\it et. al.}, Phys. Rev. B 54, 11169 (1996).

\bibitem{wannier90} A .A. Mostofi, J. R. Yates, Y. -S. Lee, I. Souza, D. Vanderbilt and N. Marzari, Comput. Phys. Commun. \textbf{178}, 685 (2008).

\bibitem{ortolani05} M. Ortolani, P. Calvani and S. Lupi, Phys. Rev. Lett. {\bf 94}, 067002 (2005).


\end{thebibliography}
\end{document}